\documentclass[epj]{svjour}

\usepackage{dcolumn}

\usepackage{amsmath}
\usepackage{amssymb}
\usepackage{graphicx}

\providecommand{\eqref}[1]{(\ref{#1})}

\newcommand{\abs}[1]{\left\vert#1\right\vert}

\newcommand{\eps}{\varepsilon}

\newcommand{\phdag}{{\phantom{\dag}}}

\newcommand{\conj}{{*}}
\newcommand{\phconj}{{\phantom{\conj}}}

\newcommand{\jprime}{{\prime}}

\newcommand{\jvecprime}{{\,\jprime}}

\newcommand{\supzero}{{(0)\!}}
\newcommand{\phsupzero}{{\phantom{(}\!}}

\newcommand{\phsubone}{{\phantom{1}}}

\newcommand{\NitrogenGroundStateTermSymbol}{\ensuremath{^1\Sigma_g^+}}
\newcommand{\NitrogenGroundState}{\ensuremath{\mathrm{N_2}(\NitrogenGroundStateTermSymbol)}}

\newcommand{\NitrogenDominantMetastableStateTermSymbol}{\ensuremath{^3\Sigma_u^+}}
\newcommand{\NitrogenDominantMetastableState}{\ensuremath{\mathrm{N_2}(\NitrogenDominantMetastableStateTermSymbol)}}

\newcommand{\NitrogenNegativeIonResonanceTermSymbol}{\ensuremath{^2\Pi_g}}
\newcommand{\NitrogenNegativeIonResonance}{\ensuremath{\mathrm{N_2^-}(\NitrogenNegativeIonResonanceTermSymbol)}}

\newcommand{\RCTCaptureME}{V_{\vec{k}}}
\newcommand{\RCTSIReleaseME}{V_{\vec{q}}}

\newcommand{\hspsi}{\hspace{0.1em}}
\newcommand{\hsdt}{\hspace{-0.6em}}
\newcommand{\hsdtexp}{\hspace{-0.15em}}

\newcommand{\Zeff}{\ensuremath Z_{\text{eff}}}

\newcommand{\AlTwoOThree}{Al$_{\mathrm{2}}$O$_{\mathrm{3}}$}
\newcommand{\MgO}{MgO}
\newcommand{\SiOTwo}{SiO$_{\mathrm{2}}$}

\begin{document}

\title{Resonant charge transfer at dielectric surfaces}

\subtitle{Electron capture and release due to impacting metastable nitrogen molecules}

\author{
	Johannes Marbach
	\and Franz Xaver Bronold
	\and Holger Fehske
}

\institute{Institut f\"ur Physik, Ernst-Moritz-Arndt-Universit\"at Greifswald, 17489 Greifswald, Germany}

\date{Received: \today / Revised version: \today}

\abstract{
We report on the theoretical description of secondary electron emission due to resonant charge transfer occurring during the 
collision of metastable~\NitrogenDominantMetastableState~molecules with dielectric surfaces. The emission is described as a 
two step process consisting of electron capture to form an intermediate shape resonance~\NitrogenNegativeIonResonance\
and subsequent electron emission by decay of this ion, either due to its natural life time or its interaction with the 
surface. The electron capture is modeled using the Keldysh Green's function technique and the negative ion decay is 
described by a combination of the Keldysh technique and a rate equation approach. We find the resonant capture of 
electrons to be very efficient and the natural decay to be clearly dominating over the surface-induced decay. 
Secondary electron emission coefficients are calculated for \AlTwoOThree, \MgO, \SiOTwo, and diamond at 
several kinetic energies of the projectile. With the exception of \MgO\ the coefficients turn out to be of the 
order of $10^{-1}$ over the whole range of kinetic energies. This rather large value is a direct consequence of 
the shape resonance acting as a relay state for electron emission. 
}

\maketitle

\section{Introduction}

Secondary electron emission from dielectric surfaces due to various projectiles is a fundamental process of great technological 
significance. The associated electron yield is usually represented by the secondary electron emission coefficient~$\gamma_e$ 
which denotes the average number of electrons released by a single collision of a particular type. Thus,~$\gamma_e$ is specific 
to the surface material, the projectile, and the particular collision process under consideration.

To illustrate the importance of secondary electron emission from dielectric materials we first consider an atmospheric 
pressure dielectric barrier discharge (DBD). Various technological processes like polymer surface treatment, thin film coating, sterilization, gas flow control and ozone production~\cite{Massines2008Recent,Golubovskii2002Influence} are based on this type of discharge. All of these are sensitive to the discharge's stability and homogeneity. Secondary electron emission from the electrodes increases the DBD's self-sustain and is thus crucial to the discharge's operation mode. In particular, there is some evidence that the stability of DBDs is controlled by secondary electron emission from plasma boundaries by impacting metastable species~\cite{Brandenburg2005Diffuse}. Although DBDs have been studied for more than 20 years, their basic mechanism is still not entirely understood~\cite{Massines2008Recent}. To overcome this deficiency numerical investigations are inevitable~\cite{Golubovskii2002Influence,Brandenburg2005Diffuse}. These studies require~$\gamma_e$ as an input parameter, which either has to be measured experimentally or calculated from first principles. 

Another application are plasma display panels (PDPs), which are used in large size TV monitors. PDPs are usually based on AC~driven high pressure microdischarges~\cite{Punset1998Twodimensional} in noble gas mixtures~\cite{Auday2000Secondary} and often employ electrodes coat\-ed with~\AlTwoOThree\ or~\MgO~\cite{Vink2002Materials}. The panel's firing voltage, that is, the voltage required for discharging, is directly influenced by secondary electron emission from the dielectric electrodes~\cite{Vink2002Materials}, for instance, due to impacting metastable species. In order to improve the performance of PDPs, theoretical investigations
of the employed materials, the overall discharge geometry, and the operation mode are necessary~\cite{Punset1998Twodimensional}. These studies again require the emission coefficient~$\gamma_e$ as an input parameter.

The demand for reliable~$\gamma_e$ data for use in numerical simulations of PDPs, DBDs, and related gas discharges
contrasts with the very sparse and often uncertain available reference data. In many cases rules of the thumb have 
to be used~\cite{Raizer91}. Especially sparse is the data base 
for secondary electron emission due to metastable molecules. The objective of the present work is therefore to 
contribute to the filling of this particular gap. We aim to establish a model that allows for easy calculation of 
the secondary electron emission coefficient due to impacting metastable molecules on dielectric surfaces, taking 
\NitrogenDominantMetastableState\ 
as an example. Other projectile-surface combinations can be modelled in the same spirit. 

\begin{figure}
        \begin{minipage}[t]{.494\linewidth}
                \begin{center}
                        \includegraphics[scale=1]{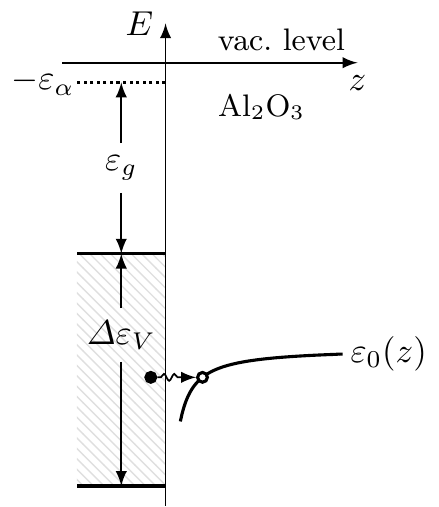}
                \end{center}
        \end{minipage}
        \begin{minipage}[t]{.494\linewidth}
                \begin{center}
                 \includegraphics[scale=1]{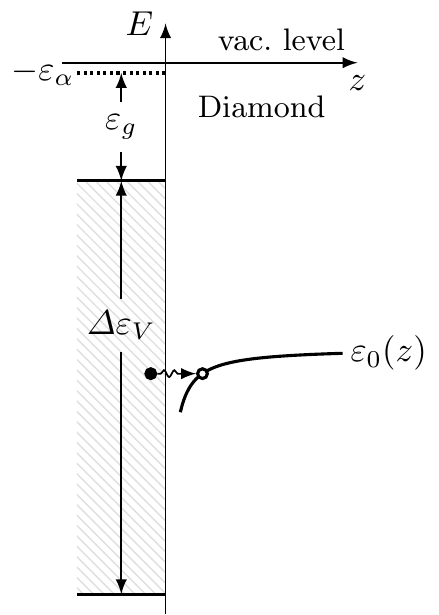}
                \end{center}
        \end{minipage}
        \begin{minipage}[t]{.494\linewidth}
                \begin{center}
                 \includegraphics[scale=1]{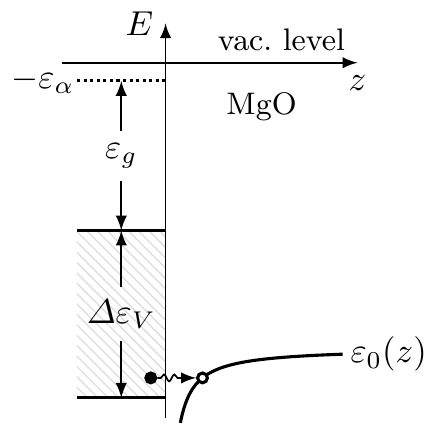}
                \end{center}
        \end{minipage}
        \begin{minipage}[t]{.494\linewidth}
                \begin{center}
                 \includegraphics[scale=1]{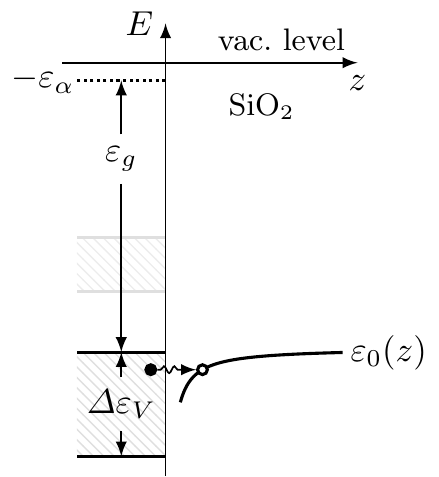}
                \end{center}
        \end{minipage}
        \caption{Energetic structure of the different dielectric materials considered in this work characterized by the electron affinity~$\eps_\alpha$, the size of the energy gap~$\eps_g$, and the width of the valence band~$\Delta\eps_V$. 
For \SiOTwo\ the energetically blocked upper valence band is marked in shaded grey. Also indicated is the variation 
of the molecular hole level with distance. For an explanation of the molecular image shift see 
Sec.~\ref{sec-capture-model}. Note that the figure is to scale in terms of energy units.}
        \label{fig-energy-schemes}
\end{figure}

The model we propose is 
an effective one, stripped of some of the microscopic details of the system and characterized by a small number of 
material specific parameters instead. In a previous work~\cite{Marbach2011Auger} we presented such a model for the 
case of Auger de-excitation of~\NitrogenDominantMetastableState\ on aluminum, corresponding to the reaction
\begin{equation}
	\NitrogenDominantMetastableState + e_{\vec{k}} \xrightarrow{\mathrm{Auger}} \NitrogenGroundState + e_{\vec{q}} \;,
	\label{eq-auger-reaction}
\end{equation}
with~$e_{\vec{k}}$ and~$e_{\vec{q}}$ denoting an electron within the surface and a free electron, respectively. 
Another important process capable of generating secondary electrons by impact of \NitrogenDominantMetastableState\ 
is the combination of electron capture by resonant charge transfer (RCT) to form the 
$\NitrogenNegativeIonResonance$~shape resonance and subsequent decay of the intermediate ion, corresponding 
to
\begin{equation}
	\NitrogenDominantMetastableState + e_{\vec{k}} \xrightarrow{\mathrm{RCT}} \NitrogenNegativeIonResonance \xrightarrow{\mathrm{RCT}} \NitrogenGroundState + e_{\vec{q}} \;.
	\label{eq-rct-reaction}
\end{equation}
Except for diamond, the Auger channel~\eqref{eq-auger-reaction} is energetically suppressed for all of the dielectric materials considered in this work whereas the RCT channel~\eqref{eq-rct-reaction} is energetically always possible 
(see Fig.~\ref{fig-energy-schemes}). Hence, we focus on the resonant 
reaction~\eqref{eq-rct-reaction} which was also found to be very efficient when 
\NitrogenDominantMetastableState\ collides with a metallic surface~\cite{Lorente1999N2,Stracke1998Formation}. In order 
to describe this process, we will use a generalization of the model presented in~\cite{Marbach2011Auger} and apply it to dielectric surfaces. Since the adaption is straightforward we only give a brief sketch of the model and refer the reader to~\cite{Marbach2011Auger} for details.

The solid surface is treated as a potential well with a depth
\begin{equation}
	V_0 = -\eps_\alpha - \eps_g - \Delta\eps_V \;,
\end{equation}
filled up to the top of the valence band
\begin{equation}
	\eps_V = -\eps_\alpha - \eps_g \;.
\end{equation}
Here~$\eps_\alpha$, $\eps_g$ and $\Delta\eps_V$ denote the solid's electron affinity, its energy gap and the width 
of its valence band, respectively. The wave functions of such a potential well are easily calculated and may be 
found in~\cite{Marbach2011Auger}. The particular electronic parameters of the dielectric materials investigated in this 
work~\cite{Yeo2002Metaldielectric,Ciraci1983Electronic,Robertson2006High,Rutter1998Ab,Robertson2006Band,Bhagavantam1948Dielectric,Kim2008Electron,Tobita2008Electronic,Fontanella1974Lowfrequency,Schreiber2002Monte,Ciraci1982Surface} are summarized in Table~\ref{tab-parameters}.

The molecule is characterized by a two level system consisting of an initially empty ground state level~$0$ and an initially filled excited level~$1$. The molecular wave functions associated with these two levels are calculated from a LCAO combination of atomic nitrogen wave functions, identifying~$0$ and $1$ with the~$2\pi_u$ and~$2\pi_g$ molecular orbitals~\cite{Marbach2011Auger}. The ionization energies $\eps_{0/1}^\infty$ of the two 
levels~\cite{Kaldor1984Generalmodelspace,Honigmann2006Complex} are listed in Table~\ref{tab-mol-params} for the unperturbed molecule. As will be shown in Sec.~\ref{sec-capture-model}, both of these levels shift downwards upon approaching the surface, provided that only the RCT channel~\eqref{eq-rct-reaction} is considered.

We will describe the resonant reaction~\eqref{eq-rct-reaction} as a two-step process. First, we will employ the Keldysh Green's function technique to model the resonant electron capture into the initially vacant ground state level~$0$. This step of the process is depicted in Fig.~\ref{fig-energy-schemes}. The second step consists of resonant electron emission by decay of the intermediate negative ion. The decay can be due to the ion-surface coupling, which is mediated by the image 
interaction, or due to the ions natural life time, that is, due to auto-decay. Considering a \NitrogenNegativeIonResonance~ion colliding with the surface we will model the surface-induced decay again by a Keldysh approach, the auto-decay, however, by a rate equation approach. Electron capture and release we will be coupled by a rate equation scheme that is based on a conditional decay reaction.

The outline of the remainder of this article is as follows. In Sec.~\ref{sec-capture} we will setup the quantum mechanical model of the molecular electron-capture and conduct a quantum-kinetic calculation on top of it. The treatment of surface-induced decay and auto-decay as well as their coupling to the initial electron capture will be presented in Section~\ref{sec-decay}. Finally, we show and discuss numerical results for selected surfaces in Sec.~\ref{sec-results} and give a conclusion in Sec.~\ref{sec-conclusion}.

\section{Resonant electron capture\label{sec-capture}}

\begin{table}
\begin{center}
  \begin{tabular}{c|c|c|c|c}
    & $\pmb{\eps_\alpha\,\,[eV]}$ & $\pmb{\eps_g\,\,[eV]}$ & $\pmb{\Delta\eps_V\,\,[eV]}$ & $\pmb{\eps_r^b}$ \\\hline
    & & & & \\[-1.5ex]
    \AlTwoOThree & 1.0 \cite{Yeo2002Metaldielectric} & 8.7 \cite{Ciraci1983Electronic} & 11.8 \cite{Ciraci1983Electronic} & 9.0 \cite{Robertson2006High} \\[1ex]
    diamond &
    $0.52$ \cite{Rutter1998Ab} & 5.48 \cite{Robertson2006Band} & 21.0 \cite{Robertson2006Band} & 5.66 \cite{Bhagavantam1948Dielectric} \\[1ex]
    \MgO & 
    $0.925$ \cite{Kim2008Electron} & 7.6 \cite{Tobita2008Electronic}& 8.5 \cite{Tobita2008Electronic} & 9.83 \cite{Fontanella1974Lowfrequency} \\[1ex]
    \SiOTwo & $0.9$~\cite{Schreiber2002Monte} & $13.8$~\cite{Ciraci1982Surface} & $5.3$~\cite{Ciraci1982Surface} & $3.9$~\cite{Robertson2006High}
  \end{tabular}
  \caption{Electronic parameters for the dielectric materials considered in this work: electron affinity~$\eps_\alpha$, energy gap~$\eps_g$, valence band width~$\Delta\eps_V$, and static relative dielectric constant~$\eps_r^b$. The electron affinity of diamond and \MgO\ is calculated from a mean value due to lack of consistent data. Note, for \SiOTwo\ only the lower valence band is considered. The upper valence band starts at~$8.9\,eV$ and has a width of~$2.7\,eV$~\cite{Ciraci1982Surface}. It
is thus energetically well separated from the molecular vacancy level~$\eps_0$ and does therefore not contribute to the electron capture.}
  \label{tab-parameters}
\end{center}
\end{table}

\subsection{Model\label{sec-capture-model}}

The molecule-surface system is described by the following Hamiltonian
\begin{equation}
\begin{split}
  H(t) & = \sum_{\vec k} \eps_{\vec k}^{\phdag} \, c_{\vec{k}}^{\dag} \, c_{\vec{k}}^{\phdag} + \eps_{0}^{\phdag}(t) \, c_{0}^{\dag} \, c_{0}^{\phdag} \\
  & \phantom{=} + \sum_{\vec k} \left( \RCTCaptureME^{\phconj\!}(t) \, c_{\vec{k}}^{\dag} \, c_{0}^{\phdag} + \RCTCaptureME^\conj(t) \, c_{\vec{k}}^{\phdag} \, c_{0}^{\dag} \right) \;,
\end{split}\label{eq-capture-hamiltonian}%
\end{equation}
where~$\vec{k}$ denotes electronic states within the solid's valence band and~$0$ labels the molecule's ground state level, which is initially empty. Since we are considering the capture of an additional electron by the metastable state~$\NitrogenDominantMetastableState$, the level~$0$ equals the lower ionization level of~$\NitrogenNegativeIonResonance$. 

To understand the variation of~$\eps_0$ with time we follow~\cite{Newns1983Charge} and consider a negative nitrogen ion 
situated at a distance~$z_R$ from a dielectric surface, the relative static permittivity of which we denote
by~$\eps_r^b$. Since the ion is negatively charged its overall energy is lowered by the value of the image potential 
at~$z_R$, for which we use the classical expression (see however below)
\begin{equation}
  V_i(z_R) \approx -\frac{\eps_r^b - 1}{\eps_r^b + 1} \frac{e^2}{4\pi\eps_0} \frac{1}{4 z_R} \;.
  \label{eq-image-potential}
\end{equation}

In order to obtain the effective ionization energy~$\eps_0(z_R)$, we virtually move the negative ion from the position~$z_R$ to a point infinitely far away from the surface, ionize the electron bound in the~$0$ state to the vacuum level, and relocate the resulting neutral molecule back to the original distance~$z_R$. The net energy change during this process determines the effective ionization energy
\begin{subequations}
\begin{align}
  \eps_0(z_R) & = -\Delta\eps = -\bigl[\; \underbrace{-V_i(z_R)}_{N_2^- \rightarrow \infty} + \underbrace{(-\eps_0^\infty)}_{N_2^- \rightarrow N_2} + \underbrace{0}_{N_2 \rightarrow z_R} \bigr] \\
  & = \eps_0^\infty + V_i(z_R) \;.
  \label{eq-eps0-zr}
\end{align}
\end{subequations}

Since this procedure is independent of the specific molecular level, we conclude that upon approaching the surface each and any of the negative ion's ionization levels shifts downwards. The reader may note that the decrease of the ionization energies does not correspond to an actual energy change of the negative ion. The ionization levels are defined solely in terms of the removal of an electron from the molecule and, thus, do not imply an energy change for a bound electron. 

Employing the usual trajectory approximation~\cite{Hagstrum1954Theory} together with the trajectory
\begin{equation}
  \vec{R}(t) = z_R(t) \, \vec{e}_z = \left( v_0 \abs{t} + z_0 \right) \vec{e}_z\;,
\end{equation}
the relation~\eqref{eq-eps0-zr} can also be understood as~$\eps_0(t)$.

\begin{table}
\begin{center}
        \begin{tabular}{c|c|c|c}
                & $\pmb{\NitrogenGroundState\,\,[eV]}$ & $\pmb{\NitrogenDominantMetastableState\,\,[eV]}$ & $\pmb{\NitrogenNegativeIonResonance\,\,[eV]}$ \\\hline
    & & & \\[-1.5ex]
                $\eps_0^\infty$ & $-17.25$~\cite{Kaldor1984Generalmodelspace} & -- & $-14.49$~\cite{Honigmann2006Complex} \\[1ex]
                $\eps_1^\infty$ & -- & $-9.67$~\cite{Kaldor1984Generalmodelspace} & $1.18$~\cite{Honigmann2006Complex} \\[1ex]
        \end{tabular}
        \caption{Ionization energies of the molecular ground state level~$0$ and the excited level~$1$ for a free nitrogen molecule in its ground state~\NitrogenGroundState\ and its metastable state~\NitrogenDominantMetastableState\ as well as for the negative ion shape resonance~\NitrogenNegativeIonResonance.}
        \label{tab-mol-params}
\end{center}
\end{table}

Generalizing Gadzuk's~\cite{Gadzuk1967Theory1} treatment of 
mono-nuclear projectiles to bi-nuclear projectiles, the tunnelling matrix element $\RCTCaptureME$ is given by 
\begin{equation}
\begin{split}
  \RCTCaptureME(t) = & \int \hspace{-0.3em} \mathrm{d}\vec{r} \; \Psi_{\vec{k}}^{\conj}(\vec{r}\hspsi) \bigl[ V_M(\vec{r}, t) + V_{SM}(z) \bigr] \, \Psi_{0}^{\phconj}(\vec{r}_{\varphi}(t)) \;,
\end{split}\label{eq-capture-me}%
\end{equation}
where the time dependence arises from the molecule's motion and~$\vec{r}_{\varphi}$ equals the vector~$\vec{r}$ as seen 
from the molecular reference frame, that is,
\begin{equation}
  \vec{r}_{\varphi}(t) = \hat\Omega(\varphi)\left[\vec{r} - \vec{R}(t)\right] \;,
  \label{eq-rphi-definition}
\end{equation}
with an appropriate rotation matrix $\hat\Omega(\varphi)$~\cite{Marbach2011Auger}. 

The expression~$V_M$ specifies the Coulomb potential of the screened molecular nuclei~\cite{Gadzuk1967Theory1}
\begin{equation}
\begin{split}
  V_M(\vec{r}, t) = & - \frac{\Zeff \, e^2}{4\pi\eps_0 \, \eps_r(z)} \left[ \frac{1}{| \vec{r} - \vec{R}_1(t) |} + \frac{1}{| \vec{r} - \vec{R}_2(t) |} \right] \;,
\end{split}%
\end{equation}
with~$\Zeff\approx4$~\cite{Marbach2011Auger} and
\begin{equation}
  \vec{R}_{1/2}(t) = \vec{R} \pm \frac{\vec{\varrho}}{2} 
\end{equation}
denoting the positions of the two nuclei, which are given 
in terms of the center of mass coordinate~$\vec{R}$ and the inter-nuclear vector~\vec{\varrho}. 
The static dielectric constant~$\eps_r(z)$ is approximated by
\begin{equation}
  \eps_r(z) = \begin{cases}
    \eps_r^b & , \, z \leq 0 \\
    1 & , \, z > 0
  \end{cases} \;.
\end{equation}

The term~$V_{SM}$ labels the electron's self image interaction potential~\cite{Gadzuk1967Theory1}
\begin{equation}
\begin{split}
  V_{SM}(z) = & \; \Theta(z) \bigl[ \Theta(z_c - z) V_0 + \Theta(z-z_c) V_i(z) \bigr] \;. \label{eq-vsm-definition}
\end{split}%
\end{equation}
Here we have used the usual truncation of the classical image potential close to the surface~\cite{Kurpick1996Basic} with
\begin{equation}
  V_i(z_c) \stackrel{!}{=} V_0 \;.
\end{equation}

The wave functions of the metal electron~$\Psi_{\vec{k}}$ and the one of the molecular electron~$\Psi_{0}$ are taken from~\cite{Marbach2011Auger}. We treat the magnetic quantum number~$m$, which is introduced within the LCAO representation of the molecular wave functions~\cite{Marbach2011Auger}, as an external parameter.

\subsection{Quantum kinetics\label{sec-quantum-kinetics-capture}}

The Hamiltonian~\eqref{eq-capture-hamiltonian} describes a standard problem of many particle physics. It can be treated 
exactly using the Keldysh technique~\cite{Blandin1976Localized}. The self-energy solely consists of a first order retarded and advanced term
\begin{subequations}
\begin{align}
  \Sigma_{\vec{k}0}^{R/A}(t_1, t_2) & = \frac{i}{\hbar} \RCTCaptureME(t_1) \, \delta(t_1 - t_2) \\
  & = -\left[\Sigma_{0\vec{k}}^{R/A}(t_1, t_2)\right]^\conj \; .
  \label{SelfenergyRA}
\end{align}%
\end{subequations}

The Keldysh component of the self-energy vanishes, the Dyson equation for the Keldysh Green's function can thus 
be solved to give
\begin{equation}
\begin{split}
  G_{\alpha\beta}^{K}(t,t^\prime) & = -i \sum_{\gamma} \bigl[ 1 - 2 n_\gamma(t_0) \bigr] \\
  & \phantom{=} \times G_{\alpha\gamma}^{R}(t,t_0) \, G_{\gamma\beta}^{A}(t_0,t^\prime) \;,
\end{split}\label{eq-gk-general}%
\end{equation}
where~$\alpha$ and~$\beta$ can take the values~$0$ or~$\vec{k}$ and~$\gamma$ runs over all states~$\{0, \vec{k}\}$ of the system. Note that~$t_0$ is shorthand for~$-\infty$. Equation~\eqref{eq-gk-general} can be used to calculate the occupation of the molecular level
\begin{subequations}
\begin{align}
  n_{0}(t) & = \frac{1}{2} \left[ 1 - i G_{00}^K(t,t) \right] \\
  & = \sum_{\vec{k}} n_{\vec{k}}(t_0) \, |G_{0\vec{k}}^R(t,t_0)|^2 \;, \label{eq-n0-general}
\end{align}%
\end{subequations}
where we have used the relation~\cite{Blandin1976Localized,Mii1996Interpolation}
\begin{equation}
  \sum_\gamma G_{\alpha\gamma}^R(t,t_0) \, G_{\gamma\alpha}^A(t_0,t) = 1 \;.
\end{equation}

Evaluation of the Green's function~$G_{0\vec{k}}^R$ appearing in~\eqref{eq-n0-general} requires knowledge of the diagonal component~$G_{00}^R$, the Dyson equation of which can be solved iteratively to yield
\begin{equation}
\begin{split}
  & G_{00}^{R\phsupzero}(t,t^\prime) = G_{00}^{R\supzero}(t,t^\prime) \, \sigma_{0}(t,t^\prime) \;,
\end{split}\label{eq-gr0-general}
\end{equation}
with
\begin{equation}
  G_{00}^{R\supzero}(t,t^\prime) = -i \Theta(t-t^\prime) \, e^{-\frac{i}{\hbar} \int_{t^\prime}^{t_\phsubone} \hsdtexp\mathrm{d}t_1 \; \eps_0(t_1)} \;,
\end{equation}
and~$\sigma_{0}$ given by the infinite series
\begin{equation}
\begin{split}
  & \sigma_{0}(t,t^\prime) = 1 - \int_{t^\prime}^{t_\phsubone} \hsdt \mathrm{d}t_1 \int_{t^\prime}^{t_1} \hsdt \mathrm{d}t_2 \; \Delta_{0}(t_1,t_2) \\
  & \quad + \int_{t^\prime}^{t_\phsubone} \hsdt \mathrm{d}t_1 \int_{t^\prime}^{t_1} \hsdt \mathrm{d}t_2 \int_{t^\prime}^{t_2} \hsdt \mathrm{d}t_3 \int_{t^\prime}^{t_3} \hsdt \mathrm{d}t_4 \; \Delta_{0}(t_1,t_2) \, \Delta_{0}(t_3,t_4) \\
  & \quad - \dots \;.
\end{split}\label{eq-sigma0-general}
\end{equation}
The quantity~$\Delta_{0}$ emerges from the self-energy and reads
\begin{equation}
\begin{split}
  \Delta_{0}(t_1,t_2) & = \frac{1}{\hbar^2} \sum_{\vec{k}} \RCTCaptureME^\conj(t_1) \, \RCTCaptureME(t_2) \\
  & \phantom{=} \times e^{-\frac{i}{\hbar} \eps_{\vec{k}} (t_1 - t_2)} \, e^{-\frac{i}{\hbar} \int_{t_1}^{t_2}\!\mathrm{d}\bar{t} \, \eps_{0}(\bar{t})} \;.
\end{split}\label{eq-delta0-definition}%
\end{equation}

By employing now the Dyson equation for~$G_{0\vec{k}}^R$, with the self energy~\eqref{SelfenergyRA}, 
the occupancy of the molecular level \eqref{eq-n0-general} can be written as 
\begin{equation}
\begin{split}
  n_{0}(t) & = \int_{t_0}^{t_\phsubone} \hsdt\mathrm{d}t_1 \int_{t_0}^{t_\phsubone} \hsdt\mathrm{d}t_2 \; \widetilde{\Delta}_{0}(t_1,t_2) \, \sigma_{0}(t,t_1) \, \sigma_{0}^\conj(t,t_2) \;.
\end{split}\label{eq-n0-exact}
\end{equation}
Here~$\widetilde{\Delta}_{0}$ is equal to~$\Delta_{0}$ with an additional factor~$n_{\vec{k}}(t_0)$ under the~$\vec{k}$ sum. In general this subtlety would deny further simplification of Eq.~\eqref{eq-n0-exact}. However, since for a dielectric surface the active energy band is the valence band, which is completely filled, the terms~$\Delta_{0}$ and~$\widetilde{\Delta}_{0}$ are essentially the same. Note that even for metallic surfaces with an incompletely filled conduction band these terms will still be almost equal, as long as the molecular resonance stays inside the occupied portion of the band. 

To simplify Eq.~\eqref{eq-n0-exact}, we first note that when~$t$ is kept constant the derivative of Eq.~\eqref{eq-sigma0-general} with respect to~$t^\prime$ can be shown to read
\begin{equation}
  \frac{\partial\sigma_{0}(t,t^\prime)}{\partial t^\prime} = \int_{t^\prime}^{t_\phsubone} \hsdt\mathrm{d}\bar{t} \; \sigma_{0}(t,\bar{t}) \, \Delta_{0}(\bar{t},t^\prime) \;.
\end{equation}
Using this expression we can now rewrite~\eqref{eq-n0-exact} into
\begin{subequations}
\begin{align}
  n_0(t) & = 2 \Re \left[ \int_{t_0}^{t_\phsubone} \hsdt\mathrm{d}t_2 \; \sigma_{0}^\conj(t,t_2) \frac{\mathrm{d} \sigma_{0}(t,t_2)}{\mathrm{d} t_2} \right] \label{eq-n0-transformation-2} \\
  \begin{split}
  & = 2 \bigl[ 1 - \sigma_{0}(t,t_0) \, \sigma_{0}^\conj(t,t_0) \bigr] \\
  & \phantom{=} - 2 \Re \left[ \int_{t_0}^{t_\phsubone} \hsdt\mathrm{d}t_2 \; \sigma_{0}(t,t_2) \frac{\mathrm{d} \sigma_{0}^\conj(t,t_2)}{\mathrm{d} t_2} \right] \;,
  \end{split} \label{eq-n0-transformation-3}
\end{align}%
\end{subequations}
where we have used integration by parts in the last line. Combining~\eqref{eq-n0-transformation-2} and~\eqref{eq-n0-transformation-3}, we obtain
\begin{equation}
  n_{0}(t) = 1 - \sigma_{0}(t,t_0) \, \sigma_{0}^\conj(t,t_0) \;.
  \label{eq-n0-simplified}
\end{equation}

We have not yet used any approximation to~$\sigma_{0}$. The result~\eqref{eq-n0-simplified} is thus exact. 
Since $\RCTCaptureME$ is small it is however sufficient to approximately re-sum the infinite 
series~\eqref{eq-sigma0-general}. Using the exponential re-summation technique described in~\cite{Marbach2011Auger} 
and retaining only the lowest order term we obtain
%
\begin{equation}
\begin{split}
  \sigma_{0}(t,t^\prime) \approx e^{-\int_{t^\prime}^{t_\phsubone} \hsdtexp \mathrm{d}t_1 \int_{t^\prime}^{t_1} \hsdtexp \mathrm{d}t_2 \; \Delta_{0}(t_1,t_2)} \;,
\end{split}\label{eq-sigma0-lowest}
\end{equation}
which after insertion into~\eqref{eq-n0-simplified} yields
\begin{equation}
  n_{0}(t) \approx 1 - e^{-\int_{t_0}^{t_\phsubone} \hsdtexp \mathrm{d}t_1 \int_{t_0}^{t_\phsubone} \hsdtexp \mathrm{d}t_2 \; \Delta_{0}(t_1,t_2)} \;.
  \label{eq-n0-lowest}
\end{equation}
Evaluating~\eqref{eq-n0-lowest} at $t=t_0$ we see that the initial condition~$n_0(t_0)=0$ is fulfilled. Taking the limit~$L\rightarrow\infty$ within the metallic wave function~\cite{Marbach2011Auger}, the~$\vec{k}$ sum in~\eqref{eq-delta0-definition} turns into an integral. For a numerical treatment it is convenient to exchange the~$\vec{k}$ and~$t_{1/2}$ integrations in~\eqref{eq-n0-lowest}. The integral in the exponent of~\eqref{eq-n0-lowest} then takes the form
\begin{equation}
\begin{split}
  \int_{t_0}^{t_\phsubone} \hsdt\mathrm{d}t_1 \int_{t_0}^{t_\phsubone} \hsdt\mathrm{d}t_2 \; \Delta_{0}(t_1,t_2) & = \frac{L^3}{2 \hbar^2 (2\pi)^3} \int\!\mathrm{d}\vec{k} \; \left| \Lambda_{0}(\vec{k},t) \right|^2 \;
\end{split}\label{eq-delta-time-integral}%
\end{equation}
with
\begin{equation}
\begin{split}
  \Lambda_{0}(\vec{k},t) = \int_{t_0}^{t_\phsubone} \hsdt\mathrm{d}t_1 \; \RCTCaptureME(t_1) \, e^{\frac{i}{\hbar} \eps_{\vec{k}} t_1} \, e^{\frac{i}{\hbar} \int_{0}^{t_1}\!\mathrm{d}t_2 \, \eps_{0}(t_2)} \;,
\end{split}\label{eq-gamma0-exact}
\end{equation}
where the~$L^3$ factor in~\eqref{eq-delta-time-integral} cancels with the~$L^{-3/2}$ factor in~$\RCTCaptureME$.

\subsection{Matrix element separation}

In order to make a numerical calculation of the ground state level's occupancy feasible, we need to introduce approximations to the matrix element. 

Numerical evaluation of the full matrix element~\eqref{eq-capture-me} suggests that~$V_M$ is clearly dominating over~$V_{SM}$ and that in addition the matrix element may be neglected in the half space~$z\leq0$. Making use of these findings we can separate the matrix element's time dependence by performing the coordinate transformation
\begin{equation}
  \vec{r} = \underbrace{\hat\Omega^\dag(\varphi) \, \vec{r}_{\varphi}}_{\vec{r}_{\varphi}^\jvecprime} + \vec{R}(t) \;,
  \label{eq-me-transformation}
\end{equation}
which is the inverse of~\eqref{eq-rphi-definition}. Utilizing the relation
\begin{subequations}
\begin{align}
  | \vec{r} - \vec{R}_{1/2}(t) | & = | \vec{r} - \vec{R}(t) \mp \frac{\varrho}{2} \, \hat\Omega^\dag(\varphi)\, \vec{e}_z | \\
  & = | \hat\Omega^\dag(\varphi) [ \vec{r}_{\varphi} \mp \frac{\varrho}{2} \, \vec{e}_z ] | \\
  & = | \vec{r}_{\varphi} \mp \frac{\varrho}{2} \, \vec{e}_z | \;
\end{align}%
\end{subequations}
and approximating~$\Theta(z^\prime_{\varphi} + z_R(t))$ by~$\Theta(z^\prime_{\varphi} + z_0)$ as in~\cite{Marbach2011Auger} 
we then obtain
\begin{equation}
  \RCTCaptureME(t) \approx C \, T_{k_z}^\conj e^{-\varkappa_{k_z} z_R(t)} \widetilde{V}_{\vec{k}} \;,
  \label{eq-exp-cap-me}%
\end{equation}
where $T_{k_z}$ is defined in~\cite{Marbach2011Auger}, 
\begin{subequations}
\begin{align}
  C & = \frac{m N_{k}}{\sqrt{2\pi N_{2\pi_u} \varkappa}} \frac{\Zeff \, e^2}{4\pi\eps_0} \;,{\rm and}\\
  \begin{split}
    \widetilde{V}_{\vec{k}} & = \int \hspace{-0.3em} \mathrm{d}\vec{r}_{\varphi} \; \Theta(z_{\varphi}^\prime - z_0) \, e^{-\varkappa_{k_z} z_{\varphi}^\prime} \left[ \frac{1}{r_{\varphi}^+} + \frac{1}{r_{\varphi}^-} \right] \\
    & \phantom{=} \times \left[ e^{-r_{\varphi}^+} + e^{-r_{\varphi}^-} \right] (x_{\varphi} + i m y_{\varphi}) \\
    & \phantom{=} \times e^{-i \left( k_x x_{\varphi}^\prime + k_y y_{\varphi}^\prime \right)} \;.	
  \end{split}
\end{align}%
\end{subequations}
Here we have rescaled~$\vec{r}_{\varphi}$,~$\vec{k}$ and~$t$ according to 
\begin{equation}
  \vec{r}_{\varphi} \rightarrow \frac{1}{\varkappa} \vec{r}_{\varphi} \;,\qquad \vec{k} \rightarrow \varkappa \vec{k} \;,\qquad t \rightarrow \frac{1}{\varkappa v_0} t \;,
  \label{eq-atomic-scaling}
\end{equation}
with~$\varkappa=\frac{\Zeff}{2 \, a_B}$. Furthermore, the expressions~$r_{\varphi}^{\pm}$ are shorthand notation for
\begin{equation}
  r_{\varphi}^{\pm} = \abs{\vec{r}_{\varphi} \pm \frac{\varrho}{2} \, \vec{e}_{z^\prime}} \;.
\end{equation}
Inserting~\eqref{eq-exp-cap-me} into Eq.~\eqref{eq-gamma0-exact} yields
\begin{equation}
\begin{split}
  \Lambda(\vec{k},t) & = \frac{C_m \, T_{k_z}^\conj \widetilde{V}_{\vec{k}}}{\varkappa v_0} \int_{t_0}^{t_\phsubone} \hsdt\mathrm{d}t_1 \; e^{-\varkappa_{k_z} z_R(t)} \, e^{\frac{i}{\hbar \varkappa v_0} \eps_{\vec{k}} t_1} \\
  & \phantom{=} \times e^{\frac{i}{\hbar \varkappa v_0} \int_{0}^{t_1}\!\mathrm{d}t_2 \, \eps_{0}(t_2)} \;,
\end{split}\label{eq-gamma0-separated}%
\end{equation}
which enables us to calculate the involved space and time integrals independently within a numerical simulation.

\section{Decay of the negative ion\label{sec-decay}}

\subsection{Surface-induced decay\label{sec-surface-decay}}

The surface-induced decay reaction can be described by the Hamiltonian
\begin{equation}
\begin{split}
  H(t) = & \sum_{\vec q} \eps_{\vec q}^{\phdag}(t) \, c_{\vec{q}}^{\dag} \, c_{\vec{q}}^{\phdag} + \eps_{1}^{\phdag}(t) \, c_{1}^{\dag} \, c_{1}^{\phdag} \\
  & + \sum_{\vec q} \left( \RCTSIReleaseME^{\phconj\!}(t) \, c_{\vec{q}}^{\dag} \, c_{1}^{\phdag} + \RCTSIReleaseME^\conj(t) \, c_{\vec{q}}^{\phdag} \, c_{1}^{\dag} \right) \;,
\end{split}\label{eq-si-release-hamiltonian}%
\end{equation}
which is structurally identical to the resonant capture Hamiltonian~\eqref{eq-capture-hamiltonian} with the associations~$\vec{k}\leftrightarrow\vec{q}$ and~$0\leftrightarrow 1$. 

In \eqref{eq-si-release-hamiltonian} $\eps_1$ is the upper ionization level of the negative nitrogen ion 
and~$\eps_{\vec{q}}$ denotes the energy of the free electron states, which are centered around the moving molecule. 
Both of these energy levels shift downwards upon approaching the surface, as can be shown using the analysis presented 
in Sec.~\ref{sec-capture-model}
\begin{subequations}
\begin{align}
  \eps_1(t) & = \eps_1^\infty + V_i(z_R(t)) \;, \\
  \eps_{\vec{q}}(t) & = \eps_{\vec{q}}^\infty + V_i(z_R(t)) \;.
\end{align}
\end{subequations}
This fact implies that the difference {$\eps_1(t) - \eps_{\vec{q}}(t)$} is constant in time.

The tunnelling matrix element~$\RCTSIReleaseME$ is given by
\begin{equation}
\begin{split}
  \RCTSIReleaseME(t) & = \alpha(\eps_{q_z}^\infty,t) \int \hspace{-0.3em} \mathrm{d}\vec{r} \; \Psi_{\vec{q}_{\varphi}}^{\conj}(\vec{r}_{\varphi}(t)) \, \Psi_{1}^{\phconj}(\vec{r}_{\varphi}(t)) \\
  & \phantom{=} \times \bigl[ V_S(z) + V_{SM}(z) \bigr] \;,
\end{split}\label{eq-release-me}%
\end{equation}
with~$\Psi_{1}$ as defined in~\cite{Marbach2011Auger} and~$\Psi_{\vec{q}}$ given by
\begin{equation}
  \Psi_{\vec{q}}(\vec{r}\hspsi) = \frac{1}{L^{\frac{3}{2}}} e^{i \vec{q} \cdot \vec{r}}~.
  \label{eq-psiq-plain-wave}
\end{equation}

Since the electron is emitted from the moving molecule, we have to evaluate in Eq.~\eqref{eq-release-me} the wave function
of the free electron,~$\Psi_{\vec{q}}(\vec{r})$, in terms of the molecular coordinates~$\vec{r}_{\varphi}$ and the wave vectors
~$\vec{q}_{\varphi}$ which are the electron's wave vectors as seen from the molecule's reference frame. The plane wave 
ansatz~\eqref{eq-psiq-plain-wave} is valid
in the present case, because the emitted electron leaves a neutral molecule behind. Conceptually, using  
$\Psi_{\vec{q}_{\varphi}}(\vec{r}_{\varphi})$ in Eq.~\eqref{eq-release-me} corresponds therefore to the 
two-center Coulomb wave used in~\cite{Marbach2011Auger} but with zero effective nucleus charge. 

In \eqref{eq-release-me} we introduced the term~$\alpha(\eps_{q_z}^\infty,t)$ which accounts for the trapping 
of slow electrons in the image potential close to the surface. When 
calculating the occupancies of the molecular levels, $\alpha$ must equal unity since the decay of the molecular state does not depend on the emitted electron being able to escape. On the other hand, when calculating the number of emitted electrons, $\alpha$ needs to cut off the matrix element when the overall energy of the emitted electron is smaller or equal to zero. 

The overall energy of the electron is given by the sum of the electron's perpendicular energy~$\eps_{q_z}^\infty$ and the image potential at the position of emission, which may be approximated by the molecule's center of mass coordinate~\cite{Marbach2011Auger}. We thus 
define 
\begin{equation}
	\alpha(\eps_{q_z}^\infty,t) = \begin{cases}
		1 & \!\!\!\!\!,\; \text{mol. level}\\
		\Theta\bigl( \eps_{q_z}^\infty + V_i(z_R(t)) \bigr) & \!\!\!\!\!,\; \vec{q}\;\text{states}
	\end{cases} \;.
	\label{eq-alpha-definition}
\end{equation}

The interaction potential~$V_S$ is given through
\begin{equation}
  V_S(z) = \Theta(-z) V_0 \;,
\end{equation}
and~$V_{SM}$ is defined in Eq.~\eqref{eq-vsm-definition}. Employing the transformation~\eqref{eq-me-transformation} the matrix element takes the form
\begin{equation}
\begin{split}
  \RCTSIReleaseME(t) & = \alpha(\eps_{q_z}^\infty,t) \int \hspace{-0.3em} \mathrm{d}\vec{r}_{\varphi} \; \Psi_{\vec{q}_{\varphi}}^{\conj}(\vec{r}_{\varphi}) \, \Psi_{0}^{\phconj}(\vec{r}_{\varphi}) \\
  & \phantom{=} \times \bigl[ V_S(z_{\varphi}^\prime + z_R(t)) + V_{SM}(z_{\varphi}^\prime + z_R(t)) \bigr] \;.
\end{split}\label{eq-transformed-release-me}%
\end{equation}

Focusing on the two principal molecule orientations and using the explicit form of the wave functions together with the calculus of Fourier transform Eq.~\eqref{eq-transformed-release-me} can be reduced to a single integral
\begin{equation}
\begin{split}
  \RCTSIReleaseME(t) & = \frac{\alpha(\eps_{q_z}^\infty,t)}{L^\frac{3}{2}}\int_{-\infty}^\infty \mathrm{d}z_{\varphi}^\prime \; e^{-i q_z z_{\varphi}^\prime} \; \Phi_\varphi(z_{\varphi}^\prime, \vec{Q}_\varphi) \\
  & \phantom{=} \times \bigl[ V_S(z_{\varphi}^\prime + z_R(t)) + V_{SM}(z_{\varphi}^\prime + z_R(t)) \bigr] \;.
\end{split}\label{eq-release-me-explicit}%
\end{equation}
In the perpendicular geometry~$\Phi_\varphi$ reads
\begin{equation}
\begin{split}
  & \Phi_\perp(z_\perp^\prime, \vec{Q}_\perp) = -(2 i) \sqrt{2\pi} \frac{Q}{(1+Q^2)^{\frac{5}{4}}} e^{i m \Phi_\perp} \\
  & \qquad \times \biggl[ \abs{z_+}^{\frac{5}{2}} K_{\frac{5}{2}}\!\left( \abs{z_+} \sqrt{1 + Q^2} \right) \\
  & \qquad - \abs{z_-}^{\frac{5}{2}} K_{\frac{5}{2}}\!\left( \abs{z_-} \sqrt{1 + Q^2} \right) \biggr] \;,
\end{split}
\end{equation}
with $K$ denoting the modified Bessel function of the second kind and
\begin{equation}
  z_\pm = z_\perp^\prime \pm \frac{\varrho}{2} \;.
\end{equation}
For the parallel orientation~$\Phi_\varphi$ takes the form
\begin{equation}
\begin{split}
  & \Phi_\parallel(z_\parallel^\prime, \vec{Q}_\parallel) = 4 i \sqrt{2\pi} \sin\left( q_x \frac{\varrho}{2} \right) \\
  & \qquad \times \biggl[ m \frac{Q \cos\Phi_\parallel}{(1+Q^2)^{\frac{5}{4}}} |z_\parallel^\prime|^{\frac{5}{2}} K_{\frac{5}{2}}\!\left( |z_\parallel^\prime| \sqrt{1 + Q^2} \right) \\
  & \qquad - z_\parallel^\prime \frac{|z_\parallel^\prime|^{\frac{3}{2}}}{(1 + Q^2)^{\frac{3}{4}}} K_{\frac{3}{2}}\!\left( |z_\parallel^\prime| \sqrt{1 + Q^2} \right) \biggr] \;,
\end{split}
\end{equation}
The lateral wave vectors~$\vec{Q}_\perp$ and~$\vec{Q}_\parallel$ are given by
\begin{subequations}
\begin{align}
  \vec{Q}_\perp & = (q_x, q_y) = (Q \cos(\Phi_\perp), Q \sin(\Phi_\perp)) \;, \\
  \vec{Q}_\parallel & = (q_y, q_x) = (Q \cos(\Phi_\parallel), Q \sin(\Phi_\parallel)) \;.
\end{align}
\end{subequations}

Since the Hamiltonian~\eqref{eq-si-release-hamiltonian} is structurally identical to~\eqref{eq-capture-hamiltonian}, the quantum-kinetic calculation outlined in Sec.~\ref{sec-quantum-kinetics-capture} holds for the current case as well. The only difference is that the ion's motion does not start at~{$t=t_0=-\infty$}, but at some later time~$t_g$, at which the ion is generated by electron capture. This can, however, easily be accounted for by replacing~$t_0$ with~$t_g$ in the previous calculation. 

Using an adapted version of Eq.~\eqref{eq-n0-general} we thus find for the occupancy of a free electron state with wave vector
$\vec{q}$
\begin{equation}
\begin{split}
  n_{\vec{q}}(t) & = \frac{1}{2} \left[ 1 - i G_{\vec{q}\vec{q}}^K(t,t) \right] = |G_{\vec{q}1}^R(t,t_g)|^2 \;.
\end{split}\label{eq-nq-general}%
\end{equation}

Employing the Dyson equation for~$G_{\vec{q}1}^R(t,t_g)$ and the results of Sec.~\ref{sec-quantum-kinetics-capture} 
the total number of emitted electrons is then given by
\begin{subequations}
\begin{align}
  n(t) & = \sum_{\vec{q}} n_{\vec{q}}(t) \\
  & = \int_{t_g}^{t_\phsubone} \hsdt\mathrm{d}t_1 \int_{t_g}^{t_\phsubone} \hsdt\mathrm{d}t_2 \; \Delta_{1}(t_1,t_2) \, \sigma_{1}^\conj(t_1,t_g) \, \sigma_{1}(t_2,t_g) \;, \label{eq-ntot-plain}\\
  & = 1 - \sigma_{1}(t,t_g) \, \sigma_{1}^\conj(t,t_g) \label{eq-ntot-simplified}
\end{align}\label{eq-ntot}
\end{subequations}
with
\begin{equation}
\begin{split}
  \Delta_{1}(t_1,t_2) & = \frac{1}{\hbar^2} \sum_{\vec{q}} \RCTSIReleaseME(t_1) \, \RCTSIReleaseME^\conj(t_2) \\
  & \phantom{=} \times e^{\frac{i}{\hbar} (\eps_{\vec{q}}^\infty - \eps_1^\infty) (t_1 - t_2)} \;
\end{split}\label{eq-lambda1-definition}%
\end{equation}
and~$\sigma_1$ defined by an expressions analogous to Eq.~\eqref{eq-sigma0-general}, but with~$\Delta_1$ instead of~$\Delta_0$. 

The equality of~\eqref{eq-ntot-plain} and~\eqref{eq-ntot-simplified}, although not obvious, can easily be confirmed. To do so, we denote these two equations by~$\zeta_1(t)$ and~$\zeta_2(t)$, respectively, and calculate their derivatives with respect to~$t$
\begin{subequations}
\begin{align}
  \begin{split}
    \frac{\mathrm{d}\zeta_1(t)}{\mathrm{d}t} & = \int_{t_g}^{t_\phsubone} \hsdt\mathrm{d}t_2 \; \Delta_{1}(t,t_2) \, \sigma_{1}^\conj(t,t_g) \, \sigma_{1}(t_2,t_g) \\
    & \phantom{=} + \int_{t_g}^{t_\phsubone} \hsdt\mathrm{d}t_1 \; \Delta_{1}(t_1,t) \, \sigma_{1}^\conj(t_1,t_g) \, \sigma_{1}(t,t_g)
  \end{split} \\
  & = - \sigma_{1}^\conj(t,t_g) \frac{\mathrm{d} \sigma_{1}(t,t_g)}{\mathrm{d}t} - \sigma_{1}(t,t_g) \frac{\mathrm{d} \sigma_{1}^\conj(t,t_g)}{\mathrm{d}t} \\
  & = \frac{\mathrm{d}\zeta_2(t)}{\mathrm{d}t} \;.
\end{align}
\end{subequations}
In addition, by inspection of Eq.~\eqref{eq-ntot} we see that~$\zeta_1(t)$ and~$\zeta_2(t)$ both vanish at~$t=t_g$. Hence,~$\zeta_1(t)$ and~$\zeta_2(t)$ are solutions of one and the same ordinary first order differential equation with one and the same initial condition. By virtue of the uniqueness of such a solution~$\zeta_1(t)$ and~$\zeta_2(t)$ are thus identical.

The time integration of the modulated matrix element involved in~\eqref{eq-ntot-simplified}, requires evaluation of
\begin{equation}
\begin{split}
  & \frac{1}{\varkappa v} \int_{t_g}^{t_\phsubone} \hsdt\mathrm{d}t_1 \; \RCTSIReleaseME(t_1) \, e^{\frac{i}{\hbar \varkappa v} (\eps_{\vec{q}}^\infty - \eps_1^\infty) t_1} \\
  & \hspace{0.68em} = \frac{1}{\varkappa v L^\frac{3}{2}}\int_{-\infty}^\infty \mathrm{d}z_{\varphi}^\prime \; e^{-i q_z z_{\varphi}^\prime} \; \Phi_\varphi(z_{\varphi}^\prime, \vec{Q}_\varphi) \, \mathcal{I}(t,t_g,z_\varphi) \;, \!\!
\end{split}\label{eq-rctr-time-integral}
\end{equation}
where we have used the variable scaling~\eqref{eq-atomic-scaling} and introduced the quantity
\begin{equation}
\begin{split}
  & \mathcal{I}(t,t_g,z_\varphi) = \int_{t_g}^{t_\phsubone} \hsdt\mathrm{d}t_1 \; \alpha(\eps_{q_z}^\infty,t_1) \, e^{\frac{i}{\hbar \varkappa v} (\eps_{\vec{q}}^\infty - \eps_1^\infty) t_1} \\
  & \qquad \times \bigl[ V_S(z_{\varphi}^\prime + z_R(t_1)) + V_{SM}(z_{\varphi}^\prime + z_R(t_1)) \bigr] \;.
\end{split}\label{eq-analytic-time-integral}
\end{equation}
Although tedious, the time integration within Eq.~\eqref{eq-analytic-time-integral} can be done analytically. For 
shortness, we merely give the result
\begin{equation}
\begin{split}
	& \mathcal{I}(t,t_g,z_\varphi) = \frac{2 V_0}{a} \biggl[ \Theta(t_{2} - t_{1}) \, \Theta(-t_g) \, e^{i a \frac{t_{1} + t_{2}}{2}} \\
	& \qquad \times \sin\!\left(a \frac{t_{2} - t_{1}}{2} \right) + \Theta(t) \, \Theta(t_{4} - t_{3}) \\
	& \qquad \times e^{i a \frac{t_{3} + t_{4}}{2}} \sin\!\left( a \frac{t_{4} - t_{3}}{2} \right)\biggr] \\
	& \qquad + \frac{\varkappa e^2}{16 \pi \eps_0} \biggl[ \Theta(t_{5} - t_g) \, \Theta(-t_g) \, e^{i a \tilde{z}} \\
	& \qquad \times \Bigl\{ \operatorname{Ei}\bigl(-i a (\tilde{z} - t_{5})\bigr) + \frac{\pi}{2} (1 + i \operatorname{sgn}(a)) \Bigr\} \\
	& \qquad - \Theta(t) \, \Theta(t - t_{6}) \, e^{- i a \tilde{z}} \\
	& \qquad \times \Bigl\{ \operatorname{Ei}\bigl(i a (\tilde{z} + t)\bigr) - \operatorname{Ei}\bigl(i a (\tilde{z} + t_{6})\bigr) \Bigr\}\biggr] \;,
\end{split}\label{eq-rctr-explicit-time-integral}%
\end{equation}
where we have introduced the following abbreviations
\begin{subequations}
\begin{align}
	t_{1} & = \max(t_g, z_\varphi^\prime + z_0 - z_c) \;, \\
	t_{2} & = \min(0, t_{q_z}, t) \;, \\
	t_{3} & = \max(t_g, t_{q_z}, 0) \;, \\
	t_{4} & = \min(z_c - z_\varphi^\prime - z_0, t) \;, \\
	\bar{t}_{5} & = z_c - z_\varphi^\prime - z_0 \;, \\
	t_{5} & = \min(-\bar{t}_{5}, -t_{q_z}, 0, t) \;, \\
	t_{6} & = \max(t_g, 0, \bar{t}_{5}, t_{q_z}) \;, \\
	t_{q_z} & = \frac{\eps_r^b - 1}{\eps_r^b + 1} \frac{\varkappa e^2}{16 \pi \eps_0} \frac{1}{\eps_{q_z}} - z_0 \;, \\
	a & = \frac{\eps_{\vec{q}}^\infty - \eps_1^\infty}{\hbar \varkappa v} \;, \\
	\tilde{z} & = z_\varphi^\prime + z_0 \;.
\end{align}\label{eq-rctr-explicit-time-integral-abbreviations}%
\end{subequations}
Equations~\eqref{eq-rctr-explicit-time-integral} and~\eqref{eq-rctr-explicit-time-integral-abbreviations} hold when calculating the number of emitted electrons $n(t)$. They can, however, also be applied to calculate the occupancy of the 
molecular levels if one sets~$t_{q_z}$ equal to~$0$. 

In applying the analysis of Sec.~\ref{sec-quantum-kinetics-capture} to the excited molecular level, we find its occupancy to be
\begin{equation}
	n_1(t) = 1 - n(t) = \sigma_{1}(t,t_g) \, \sigma_{1}^\conj(t,t_g) \;.
	\label{eq-n1}
\end{equation}

In order to obtain the rate of the surface-induced decay process~$\Gamma_S$, we use the lowest order approximation 
for~$\sigma_{1}$ and take the derivative of Eq.~\eqref{eq-n1} with respect to~$t$,
%
\begin{equation}
\begin{split}
  \frac{\mathrm{d} n_1(t)}{\mathrm{d}t} & = 2 \Re \left[ \sigma_{1}^\conj(t,t_g) \frac{\mathrm{d} \sigma_{1}(t,t_g)}{\mathrm{d}t} \right] \\
  & \approx - 2 \int_{t_g}^{t_\phsubone} \hsdt\mathrm{d}t_2 \; \Re\left[ \Delta_{1}(t,t_2) \right]\sigma_{1}(t,t_g) \sigma_{1}^\conj(t,t_g)\nonumber\\
& \equiv \Gamma_S(t,t_g)n_1(t) \;
\end{split}\label{eq-rctr-rate-definition}%
\end{equation}
with 
\begin{eqnarray}
\Gamma_S(t,t_g)=- 2 \int_{t_g}^{t_\phsubone} \hsdt\mathrm{d}t_2 \; \Re\left[ \Delta_{1}(t,t_2) \right] \;.
\end{eqnarray}
Note that in addition to the time of observation~$t$, the decay rate~$\Gamma_S$ also depends on the time of generation~$t_g$. 
The coupling of the surface-induced decay process with the prior negative ion generation will be outlined in 
Sec.~\ref{sec-consecutive-decay}. Before, however, we turn our attention to the natural decay of the shape 
resonance \NitrogenNegativeIonResonance .

\subsection{Natural decay\label{sec-natural-decay}}

The \NitrogenNegativeIonResonance\ shape resonance has a mean life time due to natural decay of~$\tau = 1.6\cdot 10^{-15}\,s$~\cite{Dreuw1999Longlived}, which corresponds to a decay rate of~$\Gamma_N = \tau^{-1} = 0.625 \cdot 10^{15} / s$. In earlier 
work~\cite{Lorente1999N2} the line shape of auto-decay processes has been modelled by means of a Lorentzian distribution, also 
known as Breit-Wigner distribution. In accordance with this approach we define therefore the natural decay rate's energy 
spectrum~$\varrho_N(\eps_{\vec{q}}^\infty,t)$ by
\begin{equation}
\begin{split}
  \Gamma_N(t) & = \int_{0}^\infty \hsdt\mathrm{d}\eps_{\vec{q}}^\infty \varrho_N(\eps_{\vec{q}}^\infty,t) \\
  & = \frac{\Delta\eps}{4\pi^3\tau} \int_{0}^\infty \hsdt\mathrm{d}\eps_{\vec{q}}^\infty \int_0^{\frac{\pi}{2}} \hsdt\mathrm{d}\vartheta_{\vec{q}} \int_0^{2\pi} \hsdt\mathrm{d}\varphi_{\vec{q}} \\
  & \phantom{=} \times \frac{\alpha\bigl( \eps_{\vec{q}}^\infty \cos^2(\vartheta_{\vec{q}}), t \bigr)}{( \eps_{\vec{q}}^\infty - \eps_{1} )^2 + \frac{\Delta\eps^2}{4}} \;.
\end{split}\label{eq-gamma-l-full}%
\end{equation}
Here~$\Delta\eps$ is the line's full width at half maximum, which can be estimated using Heisenberg's uncertainty relation
\begin{equation}
  \Delta\eps \gtrsim \frac{h}{\tau} \;.
\end{equation}

In~\eqref{eq-gamma-l-full} we again introduced the image potential trapping factor~$\alpha(\eps_{q_z},t)$. Using its explicit form~\eqref{eq-alpha-definition} we can carry out the angle integration in~\eqref{eq-gamma-l-full} and obtain
\begin{equation}
  \Gamma_N(t) = \frac{\Delta\eps}{2\pi \tau} \int_{\eps_\alpha}^\infty \hsdt\mathrm{d}\eps_{\vec{q}}^\infty \; \frac{\bar{\alpha}(\eps_{\vec{q}}^\infty,t)}{( \eps_{\vec{q}}^\infty - \eps_{1}^\infty )^2 + \frac{\Delta\eps^2}{4}} \;,
  \label{eq-gamman}
\end{equation}
with
\begin{equation}
	\bar{\alpha}(\eps_{\vec{q}}^\infty,t) = \begin{cases}
		\frac{1}{2} & \!\!\!\!\!,\; \text{mol. level} \\
		\frac{1}{\pi} \operatorname{acos}\biggl( \sqrt{\frac{\eps_i(z_R(t))}{\eps_{\vec{q}}^\infty}} \biggr) & \!\!\!\!\!,\;\vec{q}~\text{states}
	\end{cases} \;.
\end{equation}
The integration boundary~$\eps_\alpha$ appearing in~\eqref{eq-gamman} is given by
\begin{equation}
	\eps_\alpha = \begin{cases}
		0 & \!\!\!\!\!,\; \text{mol. level} \\
		\eps_i(z_R(t)) & \!\!\!\!\!,\;\vec{q}~\text{states}
	\end{cases} \;.
\end{equation}

The rate equations for the complete electron emission process needs to take into account not only the decay process, 
be it natural or surface-induced, but also the preceding generation of the negative ion. It is described next.

\subsection{Consecutive decay approach\label{sec-consecutive-decay}}

Because of the lack of an effective Hamiltonian for the active electrons describing simultaneously the formation 
of the negative ion due to electron capture and the decay of the negative ion with subsequent electron emission 
we cannot employ the Keldysh technique to calculate the secondary electron emission coefficient. As an alternative, 
we use rate equations with rates obtained from the Keldysh approach. 

The reaction chain consists of two consecutive processes. Considering only 
the molecular states we can write
\begin{equation}
  \NitrogenDominantMetastableState \xrightarrow{\;\;\Gamma_{0}(t)\;\;} \NitrogenNegativeIonResonance \xrightarrow{\;\;\Gamma_{1}(t)\;\;} \NitrogenGroundState \;,
  \label{eq-consecutive-decay-scheme}
\end{equation}
where $\Gamma_0(t)$ and $\Gamma_1(t)$ are the reaction rates of electron capture and release, respectively. The scheme~\eqref{eq-consecutive-decay-scheme} is easily translated into a system of ordinary differential equations
\begin{subequations}
\begin{align}
  \frac{\mathrm{d}n_*(t)}{\mathrm{d}t} & = -\Gamma_0(t) \, n_*(t) \;, \\
  \frac{\mathrm{d}n_-(t)}{\mathrm{d}t} & = -\Gamma_1(t) \, n_-(t) + \Gamma_0(t) \, n_*(t) \;, \label{eq-nminus-ode}\\
  \frac{\mathrm{d}n_g(t)}{\mathrm{d}t} & = \Gamma_1(t) \, n_-(t) \;, \label{eq-ng-ode}
\end{align}\label{eq-sub-dec-odes}%
\end{subequations}
where~$n_*$,~$n_-$ and~$n_g$ denote the respective fractions of metastable molecules, negative ions, and ground state 
molecules, which obey the relation
\begin{equation}
  n_*(t) + n_-(t) + n_g(t) \equiv 1 \quad \forall \; t \;
\end{equation}
and satisfy the initial conditions
\begin{equation}
	n_*(t_0) = 1 \;, \quad n_-(t_0) = 0 \;, \quad n_g(t_0) = 0 \;.
\end{equation}
The system~\eqref{eq-sub-dec-odes} can be solved straightforwardly to give
\begin{subequations}
\begin{align}
  n_*(t) & = e^{-\int_{t_0}^{t_\phsubone} \hsdtexp\mathrm{d}t_1 \; \Gamma_0(t_1)} \;, \label{eq-n*}\\
  n_-(t) & = -\int_{t_0}^{t_\phsubone} \hsdt\mathrm{d}t_1 \; \frac{\mathrm{d} n_*(t_1)}{\mathrm{d}t_1} \; e^{\int_{t_1}^{t_\phsubone} \hsdtexp\mathrm{d}t_2 \; \Gamma_1(t_2)} \;, \label{eq-n-}\\
  n_g(t) & = 1 - n_*(t) - n_-(t) \label{eq-ng} \;.
\end{align}%
\end{subequations}

The generation of negative ions is realized by filling the hole in the ground state level~$0$, that is, the 
lower ionization level of~$\NitrogenNegativeIonResonance$. The ion then decays by emitting an electron from the 
excited level~$1$ which is the higher ionization level of~$\NitrogenNegativeIonResonance$. Hence, the molecular 
fractions~$n_{*/-/g}$ can be related to the single level occupancies~$n_{0/1}$ by means of
\begin{equation}
  \left.
  \begin{aligned}
  n_*(t) & = 1 - n_0(t) \\
  n_g(t) & = 1 - n_1(t)
  \end{aligned}
  \right\} \; \Rightarrow \; n_-(t) = n_0(t) + n_1(t) - 1 \;.
  \label{eq-n0n1-n*ng}
\end{equation}

To calculate the total number of emitted electrons~$n(t)$ and their energy spectrum we first note that due to particle conservation~$n(t)$ must equal the fraction of ground state molecules~$n_g(t)$. Hence, we can employ Eq.~\eqref{eq-ng-ode} to calculate~$n(t)$. In analogy to \eqref{eq-gamma-l-full} we write the decay rate~$\Gamma_1(t)$ as
\begin{equation}
  \Gamma_1(t) = \int_{0}^{\infty} \hsdt\mathrm{d}\eps_{\vec{q}}^\infty \; \varrho_1(\eps_{\vec{q}}^\infty,t) \;,
  \label{eq-gamma1-energy-distribution-definition}
\end{equation}
where we introduced the rate's energy spectrum~$\varrho_1(\eps_{\vec{q}}^\infty,t)$ which needs to account 
for trapping of low energy electrons in front of the surface due to the image potential. It will be given below. 

After employing~\eqref{eq-gamma1-energy-distribution-definition} in~\eqref{eq-ng-ode} we directly integrate the 
resulting equation and obtain
\begin{equation}
  n(t) = \int_0^{\infty} \hsdt\mathrm{d}\eps_{\vec{q}}^\infty \int_{t_0}^{t_\phsubone} \hsdt\mathrm{d}t_1 \; \varrho_1(\eps_{\vec{q}}^\infty,t_1) \, n_-(t_1) \;.
  \label{eq-consecutive-spectrum}
\end{equation}
Inserting~$n_-$ from Eq.~\eqref{eq-n-} we can use~\eqref{eq-consecutive-spectrum} to calculate the overall number of emitted electrons as well as their energy spectrum. For~$t=\infty$ Eq.~\eqref{eq-consecutive-spectrum} defines the secondary electron emission coefficient~$\gamma_e=n(\infty)$.

Finally, we need to build a link between the rates~$\Gamma_{0/1}$ and the previously calculated quantities~$\Gamma_{S/N}$. 
The rate of electron capture~$\Gamma_0$ is defined by
\begin{equation}
	\frac{\mathrm{d} \bigl( 1 - n_0(t) \bigr)}{\mathrm{d} t} = - \Gamma_0(t) \, \bigl( 1 - n_0(t) \bigr) \;.
\end{equation}
and can be calculated from~\eqref{eq-n0-lowest}. The rate of negative ion decay~$\Gamma_1$ contains contributions from the surface-induced decay and the natural decay. For the latter the rate~$\Gamma_N$ was explicitly given in Sec.~\ref{sec-natural-decay}. For the former we need to keep in mind that the rate~$\Gamma_S(t,t_g)$ given in Sec.~\ref{sec-surface-decay} also depends on the time of negative ion creation~$t_g$. To obtain the overall rate~$\Gamma_S(t)$, we therefore need to weight the two-time rate~$\Gamma_S(t,t_g)$ with the negative ion's production term along the trajectory, that is,
\begin{equation}
	\Gamma_S(t) = \int_{t_0}^{t_\phsubone} \hsdt\mathrm{d}t_g \; \Gamma_S(t,t_g) \, \underbrace{\Gamma_0(t_g) \, n_*(t_g)}_{-\frac{\mathrm{d}n_*(t_g)}{\mathrm{d}t_g}} \;.
\end{equation}
Employing the explicit form of~$\Gamma_S(t,t_g)$ given by Eq.~\eqref{eq-rctr-rate-definition} together with partial integration we obtain
\begin{equation}
	\Gamma_S(t) = \int_{t_0}^{t_\phsubone} \hsdt\mathrm{d}t_1 \; 2 \Re \bigl[ \Delta_1(t,t_1) \bigr] \bigl( 1 - n_*(t_1) \bigr) \;.
	\label{eq-gammas}
\end{equation}
The total rate of decay is given by
\begin{equation}
  \Gamma_1(t) = \Gamma_S(t) + \Gamma_N(t) \;
\end{equation}
and the total rate's energy distribution~$\varrho_1(\eps_{\vec{q}}^\infty,t)$, which was introduced in~\eqref{eq-gamma1-energy-distribution-definition}, is related to the previous calculation by means of~\eqref{eq-gamman} and~\eqref{eq-gammas}, respectively. 

To be specific we note that for constructing $\varrho_1(\eps_{\vec{q}}^\infty,t)$ the $\vec{q}$~sum involved 
in~$\Delta_1$ (see Eq.~\eqref{eq-lambda1-definition}) has to be converted into an integral by letting~$L\rightarrow\infty$. 
Afterwards one can integrate the angle coordinates of the $\vec{q}$~vector to obtain the energy spectrum of the
rate for the surface-induced decay. The final expression for $\varrho_1(\eps_{\vec{q}}^\infty,t)$ reads
\begin{equation}
	\varrho_1(\eps_{\vec{q}}^\infty,t) = \varrho_S(\eps_{\vec{q}}^\infty,t) + \varrho_N(\eps_{\vec{q}}^\infty,t) \;,
\end{equation}
with
\begin{equation}
\begin{split}
	\varrho_S(\eps_{\vec{q}}^\infty,t) & = \left( \frac{2 m_e}{h^2 \varkappa^2} \right)^{\frac{3}{2}} \frac{L^3}{\hbar^2} \int_{t_0}^{t_\phsubone} \hsdt\mathrm{d}t_1 \int_0^{\frac{\pi}{2}} \hsdt\mathrm{d}\vartheta_{\vec{q}} \int_0^{2\pi} \hsdt\mathrm{d}\varphi_{\vec{q}} \\
	& \phantom{=} \times \Re\left\{\RCTSIReleaseME(t) \, \RCTSIReleaseME^\conj(t_1) \, e^{\frac{i}{\hbar} (\eps_{\vec{q}}^\infty - \eps_1^\infty) (t - t_1)} \right\} \\
	& \phantom{=} \times \sqrt{\eps_{\vec{q}}^\infty} \, \sin(\vartheta_{\vec{q}}) \bigl( 1 - n_*(t_1) \bigr) \;
\end{split}%
\end{equation}
and $\varrho_N(\eps_{\vec{q}}^\infty,t)$ implicitly defined in \eqref{eq-gamma-l-full}.

\section{Results\label{sec-results}}

\begin{figure}
  \centering
  \includegraphics[scale=1]{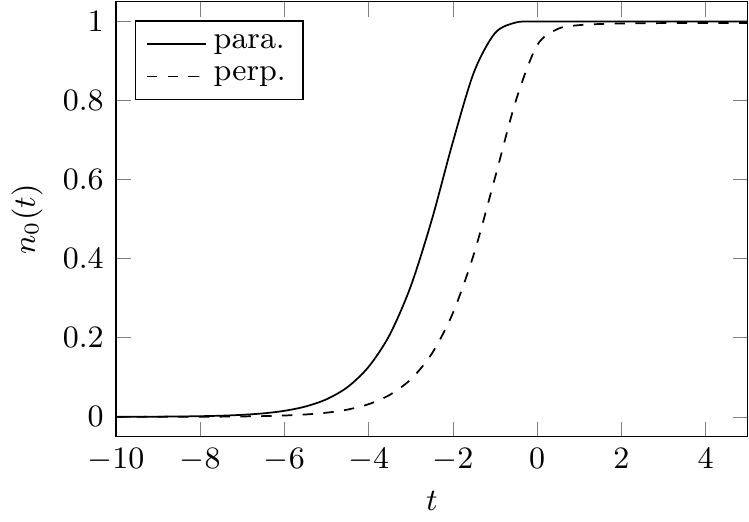}
  \caption{Time evolution of the occupation of the molecular ground state level~$n_0$ in front of an \AlTwoOThree~surface for parallel (solid line) and perpendicular (dashed line) molecule orientation. The molecule's kinetic energy was fixed to~$50\,meV$. The curves were calculated from Eq.~\eqref{eq-n0-lowest}.\label{fig-n0}}
\end{figure}

In the following we present numerical results for the two principal molecule orientations, that is, for the molecular axis aligned perpendicular ($\varphi=\frac{\pi}{2}$) and parallel ($\varphi=0$) to the surface. 

First, we consider the capturing of a metal electron into the molecular ground state level, which was described in Sec.~\ref{sec-capture}. Figure~\ref{fig-n0} shows the evolution of the ground state level's occupation~$n_0(t)$ as calculated from Eq.~\eqref{eq-n0-lowest} for the case of an \AlTwoOThree~surface at a kinetic energy of~$50\,meV$. We clearly see that the process is very efficient, since the molecular vacancy is completely filled within the incoming branch of the trajectory for both orientations.The efficiency is higher in parallel orientation. Complete filling of the ground state hole is achieved in that 
case at about~$t=-1$, corresponding to a molecule-surface distance of approximately~$4.92\,a_B$. In the perpendicular 
orientation, on the other hand, complete filling is only reached at about~$t=0$ which corresponds to the molecule's 
turning point~$z_0=4.42\,a_B$.

\begin{figure}
  \centering
  \includegraphics[scale=1]{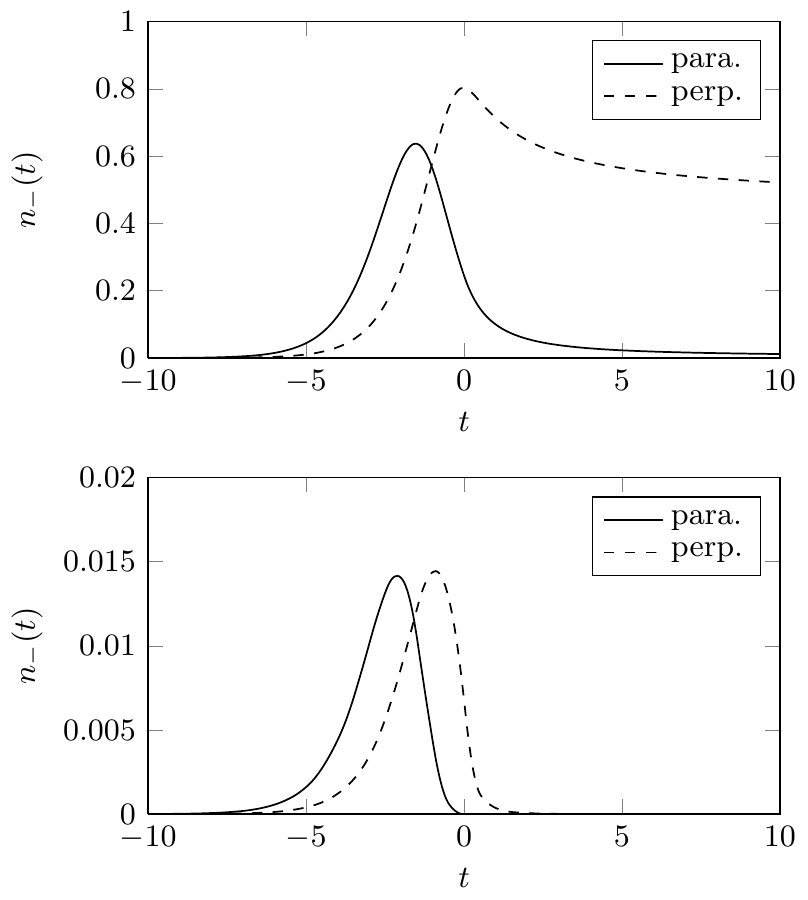}
  \caption{\label{fig-n-}Fraction of negative ions versus time for the surface-induced decay process (upper panel) and the natural decay reaction (lower panel) in front of an \AlTwoOThree~surface at a molecular kinetic energy of~$50\,meV$. The solid lines represent the parallel molecule orientation, whereas the dashed lines denote the perpendicular orientation. The results were calculated from Eq.~\eqref{eq-n-}.}
\end{figure}

Next, we focus on the decay of a negative ion generated by electron capture. To estimate the relative strength of the surface-induced and the natural decay channel, respectively, we treat them separately. Figure~\ref{fig-n-} shows the time evolution of the fraction of negative ions as calculated from Eq.~\eqref{eq-n-} for both processes in front of an \AlTwoOThree~surface at~$\eps_{\text{kin}}=50\,meV$. As we see, within the incoming branch the portion of negative ions first increases due to the efficiency of the electron capture into the ground state and the finiteness of the decay rates. At some point within the incoming branch, however, the decay process starts to outbalance the generating reaction, resulting in a decrease of the negative ion fraction. Since the surface-induced decay process is driven by the molecule-surface interaction~\eqref{eq-release-me}, which vanishes at large distances from the solid, the negative ion share saturates to a finite value in this case (see upper panel of Fig.~\ref{fig-n-}). For the parallel orientation the saturation level is much lower than for the perpendicular orientation. This is caused by the antisymmetry of the molecular wave functions in the plane normal to the molecule's axis. In the parallel orientation this antisymmetry is broken by the surface, which results in a significantly enlarged matrix element, and hence in a higher rate of decay. For the natural decay, on the other hand, the decay rate is constant, and thus the negative ion fraction eventually drops to zero (see lower panel of Fig.~\ref{fig-n-}). 

An important point to notice is that the portion of negative ions is about two orders of magnitude higher in the surface-induced decay as compared to the natural decay. The 
natural decay rate must thus be considerably larger than the surface-induced decay rate. Hence, the latter can be neglected in the present situation. It could, however, become important for material combinations that allow for projectile turning points closer to the surface. This would lead to an increase in the matrix element, which in turn results in a higher decay rate.

\begin{figure}
  \centering
  \includegraphics{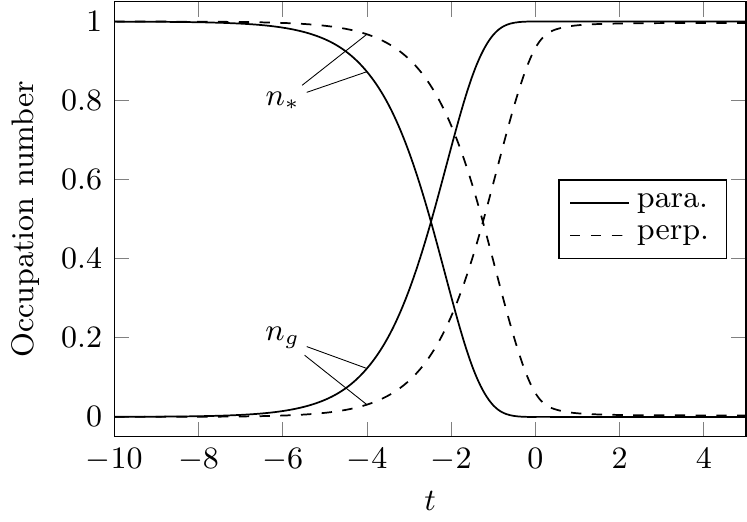}
  \caption{\label{fig-n0n1}Time evolution of the fractions of metastables~$n_*$ and ground state molecules~$n_g$ in parallel (solid lines) and perpendicular (dashed lines) orientation for an \AlTwoOThree~surface at a molecular kinetic energy of~$50\,meV$. Only the natural decay channel was considered. The curves were obtained by employing Eqs.~\eqref{eq-n*} and~\eqref{eq-ng}, respectively.}
\end{figure}

To calculate the fractions of metastables~$n_*$ and ground state molecules~$n_g$ from Eqs.~\eqref{eq-n*} and~\eqref{eq-ng}, 
respectively, we focus therefore, in view of the above discussion, exclusively on the natural decay and neglect the 
surface-induced decay channel.
The results are depicted in Fig.~\ref{fig-n0n1}. Note that with the decrease of the fraction of metastables the ground state share almost immediately rises by the same amount, which is due to the rather large decay rate for natural decay. 
The fractions~$n_*$ and~$n_g$ can be related to the occupancies of the molecular levels~$n_0$ and~$n_1$ by means of Eq.~\eqref{eq-n0n1-n*ng}.

\begin{figure}
  \centering
  \includegraphics[scale=1]{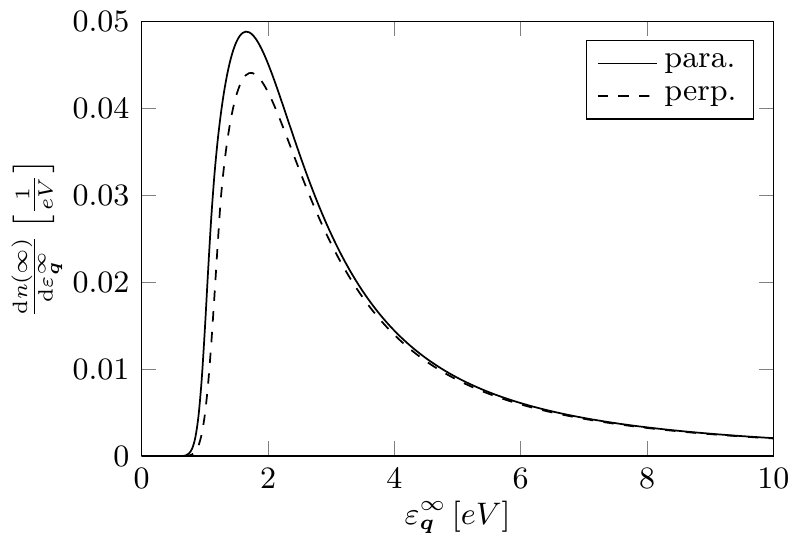}
  \caption{\label{fig-spectrum}Spectrum of emitted electrons for an \AlTwoOThree~surface in parallel (solid lines) and perpendicular (dashed lines) orientation at $t=\infty$, as calculated from Eq.~\eqref{eq-consecutive-spectrum}. The molecule's kinetic energy was fixed to~$50\,meV$. Only the natural decay process was taken into account.}
\end{figure}

We now turn to the spectrum of emitted electrons, which can be calculated from Eq.~\eqref{eq-consecutive-spectrum}. 
Figure~\ref{fig-spectrum} shows the electronic spectrum at $t=\infty$, again for the case of \AlTwoOThree\ and 
$\eps_\text{kin}=50\,meV$, while neglecting the surface-induced decay channel. In both molecular orientations the 
spectrum exhibits a strong cut-off for energies below approximately~$1\,eV$. This is a direct consequence of the 
image potential, trapping low energy electrons close to the surface, which was incorporated into our calculation 
by means of the quantity~$\alpha(\eps_{q_z},t)$ (see Eq.~\eqref{eq-alpha-definition}). Following the low energy 
cut-off, the spectra show a strong peak at about~$1.8-1.9\,eV$, before they slowly fall off towards zero for 
larger energies. Examining the curves we further note that the spectrum is larger in the parallel orientation than 
in the perpendicular orientation. This can again be attributed to the trapping effect. From Fig.~\ref{fig-n-} we 
know that in perpendicular geometry the negative ion is generated and destroyed closer to the surface than 
in parallel orientation. The image potential and hence the trapping effect are thus particularly strong in that
case which in turn reduces the efficiency of electron emission.

Finally, we calculate the secondary electron emission coefficient~$\gamma_e$, which is the area beneath 
curves of the type shown in Fig.~\ref{fig-spectrum}. We specifically consider \AlTwoOThree, \MgO, \SiOTwo, 
and diamond. The 
electronic parameters for these substances are listed in Table~\ref{tab-parameters}. We assume the turning point 
formula given in~\cite{Marbach2011Auger} to be valid for all of these materials. Hence, we can neglect the 
surface-induced decay process. 

\begin{figure}
  \centering
  \includegraphics[scale=1]{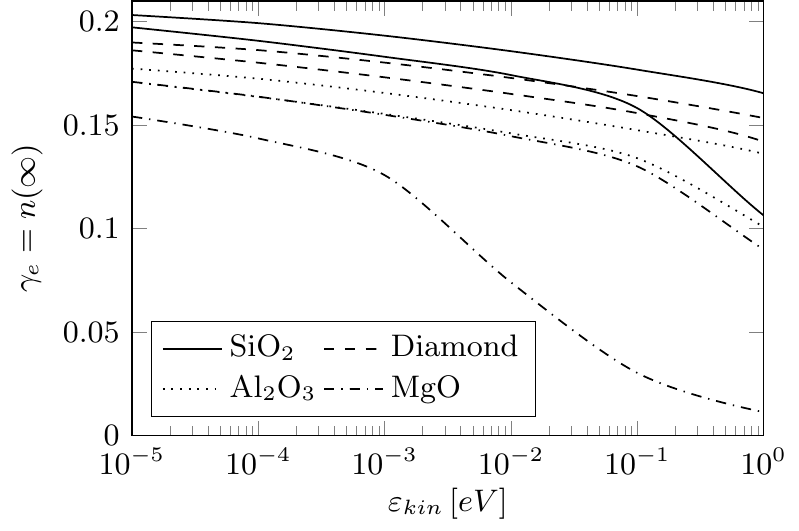}
  \caption{\label{fig-gamma}Secondary electron emission coefficient~$\gamma_e$ for several different dielectric surfaces plotted against the molecule's kinetic energy~$\eps_{kin}$. For each material two curves are depicted. The upper corresponds to the parallel molecule orientation, whereas the lower identifies the perpendicular molecule orientation. All results were obtained from Eq.~\eqref{eq-consecutive-spectrum}.}
\end{figure}

Figure~\ref{fig-gamma} depicts the variation of the secondary electron emission 
coefficient~$\gamma_e$ with the molecule's kinetic energy~$\eps_{kin}$ for the aforementioned materials as 
calculated from Eq.~\eqref{eq-consecutive-spectrum}. Leaving \MgO\ aside (for a discussion see below), 
the $\gamma_e$-coefficients are on the 
order of $10^{-1}$ over the whole range of kinetic energies. The rather large values for $\gamma_e$ 
can be attributed to the shape resonance \NitrogenNegativeIonResonance\ which is not only efficiently 
formed in front of the surface but also quickly decays thereby releasing an electron. Atomic projectiles, 
for instance, metastable 
argon, which cannot form a metastable negative ion, will lead to much smaller $\gamma_e$ values. 

The secondary electron emission coefficient we find for metastable nitrogen hitting a dielectric surface
agrees rather well with the $\gamma_e$-coefficient which is required for a convergent, self-consistent 
kinetic simulation of DBDs taking this particular surface 
collision process into account~\cite{Brandenburg2005Diffuse}. In keeping with our 
initial motivation, the effective microscopic model is thus indeed capable to provide input data with 
sufficient accuracy for the kinetic modelling of gas discharges.

Let us now analyze Fig.~\ref{fig-gamma} in more detail. The $\gamma_e$-coefficients increase with decreasing 
kinetic energy. This can be explained by the enlarged 
molecule-surface interaction time and the narrowed turning point which lead to a more effective filling of 
the ground state hole at lower projectile velocities. Apart from this obvious monotony property, the specific 
form of the particular graphs is, however, not easily understood. The reason for this is the complex 
dependence of the secondary electron emission coefficient on the surface's band structure and its static 
dielectric constant. For the materials we considered a few general remarks, however, can be given.

First, we note that the emission coefficient is sensitive to the efficiency of the electron capture into the 
molecular hole level. A high efficiency results in negative ions getting produced at large distances from the 
surface. The subsequent electron emission will thus also take place at large distances, because of the large natural 
decay rate of the ions. This is beneficial to the emission coefficient, as the image potential trapping is less 
severe at larger distances. The efficiency of the electron capture in turn is determined by the alignment of the
solid's valence band to the molecular hole level. Since the filling of the molecular vacancy is a resonant 
tunneling process, the vacancy level~$\eps_0$ has to be degenerate to occupied states within the surface in order 
to allow for an efficient generation of negative ions. In addition, the electron capture is more productive if 
the degeneracy appears at higher energies within the valence band. This is due to the fact that at higher energies 
the wave functions of the electrons within the solid have a larger extension outside the solid. Thus their overlap 
with the molecular wave function is increased, which directly influences the matrix element and thus the efficiency 
of the electron capture. For \MgO\ both conditions, degeneracy and higher energies within the valence band, are 
violated since the molecular hole level~$\eps_0$ shifts downwards out of the valence band at small distances 
(see Fig.~\ref{fig-energy-schemes}). As a result \MgO\ has the smallest emission coefficient of the materials 
under consideration. 

Secondly, a higher dielectric constant seems to have an adverse effect on the value of~$\gamma_e$, since for the most part the emission coefficients for~\SiOTwo\ and diamond, both having lower dielectric constants, are higher than the emission coefficient of~\AlTwoOThree\, which possesses a higher dielectric constant (see Table~\ref{tab-parameters}). This effect can probably be attributed to the amount of image potential trapping of slow electrons, which is proportional to~$(\eps_r^b - 1) / (\eps_r^b + 1)$ (see Eqs.~\eqref{eq-image-potential} and~\eqref{eq-alpha-definition}). This factor increases with $\eps_r^b$. Thus,
the image potential and with it the effect of trapping is enlarged for higher dielectric constants.

\section{Conclusion\label{sec-conclusion}}

In this work we described a generic effective microscopic model for secondary electron emission 
due to collision of metastable molecules with dielectric surfaces. As an illustration we applied 
the model to \NitrogenDominantMetastableState\ hitting a dielectric surface, focusing 
on the RCT channel~\eqref{eq-rct-reaction} which is particularly strong because this molecule
forms a shape resonance upon electron capture. In addition, the 
competing Auger process~\eqref{eq-auger-reaction} is energetically blocked for the 
dielectrics commonly used in gas discharge physics and needs thus not to be considered. 

Our model depends only on a small number of material parameters. As far as the surface is concerned, 
these parameters include the electron affinity~$\eps_\alpha$ of the solid, its band gap $\eps_g$, and 
the width of its valence band~$\Delta\eps_V$ as well as its static dielectric constant~$\eps_r^b$. Any 
surface that can be parameterized by such a set can in general be treated within our model. With a few 
modifications the model can also be applied to metallic surfaces and other molecular and atomic 
projectiles.

We presented numerical results for \AlTwoOThree, \MgO, \SiOTwo, and diamond surfaces. The electronic 
parameters of the surface in question are crucial for the overall efficiency of the RCT process. The 
generation of the shape resonance~\NitrogenNegativeIonResonance\ upon electron capture into the lower 
ionization level of~\NitrogenNegativeIonResonance\ is efficient only when the molecular level stays 
well inside the valence band. Concerning the decay of the temporary negative ion we found the 
surface-induced process to be about two orders of magnitude 
weaker than the auto-decay. The surface-induced decay channel can thus be 
neglected for the situations we investigated. We found the spectrum of the emitted electron to be 
strongly peaked at about~$1.8-1.9\,eV$. In a gas discharge, however, the auto-decay occurs close 
to the plasma wall where the sheath potential is large. The emitted electron will thus 
immediately gain a kinetic energy which is substantially larger than the energy resulting from
auto-decay.

Calculations of the secondary electron emission coefficient~$\gamma_e$ for different surface
materials finally yielded $\gamma_e$-values from~$0.02$ to~$0.2$ for collision energies ranging 
from~$0.1\,\mu eV$ to~$1\,eV$. Particularly for thermal energies the $\gamma_e$-values we obtain 
are of the order of $10^{-1}$ and thus coincide with the values deduced from kinetic simulations 
of dielectric barrier discharges which take this particular secondary electron emission process
into account. Hence, despite its simplicity
effective microscopic modelling can capture the essential physics of secondary electron 
emission from surfaces due to impacting metastable molecules and is thus capable to fill some 
of the gaps in the molecule-surface collisions' reference data which are required for the kinetic 
modeling of gas discharges.

\section*{Acknowledgments}

Johannes Marbach was funded by the federal state of Mecklenburg-Western Pomerania through a postgraduate scholarship. In addition this work was supported by the Deutsche Forschungsgemeinschaft through the Transregional Collaborative Research Center SFB/TRR24.

\bibliographystyle{epj}
\bibliography{main}

\end{document}